# Sample size for a non-inferiority clinical trial with time-to-event data in the presence of competing risks[§]

**Dong Han[1,2],   Zheng Chen[1]*,   Yawen Hou[3]**

[1] Department of Biostatistics, School of Public Health, Southern Medical University, Guangzhou 510515, China;

[2] Department of Quality Control, The Third Affiliated Hospital of Southern Medical University, Guangzhou 510665, China;

[3] Department of Statistics, College of Economics, Jinan University, Guangzhou 510632, China

*: Corresponding author

## Abstract

The analysis and planning methods for competing risks model have been described in the literatures in recent decades, and non-inferiority clinical trials are helpful in current pharmaceutical practice. Analytical methods for non-inferiority clinical trials in the presence of competing risks were investigated by Parpia et al., who indicated that the proportional sub-distribution hazard model is appropriate in the context of biological studies. However, the analytical methods of competing risks model differ from those appropriate for analyzing non-inferiority clinical trials with a single outcome; thus, a corresponding method for planning such trials is necessary. A sample size formula for non-inferiority clinical trials in the presence of competing risks based on the proportional sub-distribution hazard model is presented in this paper. The primary endpoint relies on the sub-distribution hazard ratio. A total of 120 simulations and an example based on a randomized controlled trial verified the empirical performance of the presented formula. The results demonstrate that the empirical power of sample size formulas based on the Weibull distribution for non-inferiority clinical trials with competing risks can reach the targeted power.

## 1. Background

Anderson et al. (2012) described the concept of generalized survival analysis, which mainly refers to multi-state model, of which single-event survival analysis and the competing risks model are special cases. Single-event survival analysis is used to analyze survival data in which an event can only occur once for each patient. In a competing risks framework, more than one type of event can occur to the patients under follow-up, and all of the events are absorbing (absorbing state). The competing event prevents the event of interest from occurring. Numerous studies have focused on methods for the inference and modeling of competing risks data in recent decades, but few designed studies that are based on a competing risks framework have been performed (Pintilie, 2002; Schulgen, 2005; Maki, 2006; Latouche, 2004; Latouche, 2007; Tai, 2015).

Sample size determination is an important aspect of clinical trial planning and design. Sample size formulas have been published for trials in the presence of competing risks (Pintilie, 2002; Schulgen, 2005; Maki, 2006; Latouche, 2004; Latouche, 2007; Tai, 2015), and the null hypothesis in all of these trials is no difference between two treatments, which is termed superiority testing. Some sample size formulas are developed based on the cause-specific hazard (CSH) ratio (Pintilie, 2002; Schulgen, 2005; Maki, 2006), whereas others use the sub-distribution hazard (SDH) ratio (Latouche, 2004; Latouche, 2007).

However, non-inferiority designs are necessary in some cases because similarity to an existing effective therapy can justify a product being on the market and providing more treatment options for patients. When competing risks are present, proportional sub-distribution hazards, Cox proportional hazards and marginal models can be used for descriptive purposes and inference. Parpia et al. (2013) discussed the performance of these methods in non-inferiority clinical trials in the presence of competing risks (NiCTCR) and indicated that aside from biological studies, the sub-distribution hazard model is preferred because of its advantage of intuitive explanation. Although non-inferiority method in clinical trials in the presence of competing risks is as necessary as method in clinical trials with single-event survival analysis, the lack of an applicable sample size determination method for planning NiCTCR has limited its development. Therefore, we present a sample size formula for NiCTCR in this paper.

The article is organized as follows. In Section 2, background information on competing risks and associated modeling methods of competing risks are introduced. A sample size formula for NiCTCR with the Weibull survival time assumption is presented. A numerical simulation is conducted to verify the usefulness of the sample size formula in non-inferiority clinical trials in the presence of competing risks in Section 3, and the results from the simulation are presented. The presented formula for NiCTCR is illustrated with an example in Section 4. Section 5 discusses various considerations of parameter settings and potential arguments. Detailed technical derivations are provided in the Appendix.

## 2. Methods

### Basic quantities of the competing risks model

More than one type of event can occur during the follow-up period in the framework of the competing risks model. In this paper, two events are considered without a loss of generality. The primary endpoint is the time to some negative event of interest, whereas a separate event is regarded as the competing event. The negative event of interest, for example, cancer death or progression of cancer, indicates that a high hazard rate corresponds to unsatisfactory efficacy. The occurrence of the competing event may affect the event of interest. Some basic descriptive statistics and models for the framework of competing risks are provided in this section.

Let $T_i$ be the recorded time to failure of the $i$th subject, where $i = 1,\ldots,N$. $N$ denotes the total sample size. $D$ implies the state code, which is the event of interest ($D = 1$), competing event ($D = 2$) or censoring ($D = 0$). $d_i$ denotes the state code of the $i$th subject.

The CSH function for the abovementioned event of interest is expressed as follows:

$$h^{CS}(t) = \lim_{\Delta t \to 0} \frac{P(t \leq T \leq t + \Delta t, D = 1 \mid T \geq t)}{\Delta t}.$$

The risk set of CSH includes subjects who are still under follow-up and who have not experienced any event, i.e., $RS^{CS}(t) = \{i \mid T_i \geq t, i = 1,\ldots,N\}$.

The other basic descriptive statistic, the cumulative incidence function (CIF), is a measure of the actual probability of events because it considers the competing cause when estimating the incidence rate for the event of interest. The CIF for the event of interest is defined as $F(t) = P(T \leq t) = \int_0^t S(u)h^{CS}(u)du$, where $S(t)$ is the event-free survival function. By derivation, $F(\infty) = P(D = 1)$ can be obtained. Therefore, the CIF is not an appropriate probability distribution. The CIF derivative $f(t) = \frac{\partial F(t)}{\partial t}$ is defined as the sub-density function.

Using the CIF as the basis for construction, the SDH for the event of interest is defined as follows:

$$h^{SD}(t) = \lim_{\Delta t \to 0} \frac{P\{t \leq T \leq t + \Delta t, D = 1 \mid T \geq t \cup (T \leq t \cap D \neq 1)\}}{\Delta t}.$$

The risk set of the SDH model is as follows:

$RS^{SD}(t) = \{i \mid T_i \geq t \cup (T_i \leq t \cap d_i \neq 1), i = 1, \ldots, N\}$.

A semi-parametric proportional hazards model for the SDH of the failure times of interest (Fine, 1999) is constructed as follows:

$$h^{SD}(t \mid x) = h_0^{SD}(t) \exp(bx),$$ (1)

where $h_0^{SD}(t)$ is the baseline sub-distribution hazard for the event of interest, and $x$ and $b$ denote the covariate and its coefficient, respectively.

Based on the definition of CSH, the sample size formula for comparing the CSHs in non-inferiority clinical trials is the same as that for comparing hazards in the Cox proportional hazard model (Formula (3) with $\theta_1 = 1$ and a truncated exponential enrollment distribution in Crisp et al. (2008)), as follows:

$$N_x^{CS} = \frac{(Z_\beta + Z_{\alpha/2})^2}{[\ln(\Delta_0^{CS})]^2 \, p_0 p_1 E^{CS}} \, p_x, \, x = 0, \, 1,$$ (2)

where $N_x^{CS}$ is the sample size required for group $x$, $E^{CS}$ is the probability of the event of interest that occurs in the entire study period after regarding the competing event as censoring, and $\Delta_0^{CS}$ is the non-inferiority margin of CSH. Type I and type II errors are denoted by $\alpha$ and $\beta$, respectively. $Z_\gamma$ denotes the $(1-\gamma)$-quartile of the standard Gaussian distribution, and $p_0$ and $p_1$ indicate the sample size percentage of groups 0 and 1, respectively.

Based on the abovementioned definitions and previous studies on the competing risks model (Lau, 2009; Putter, 2007), the CSH censors the competing events, but such censoring is assumed to be non-informative because the censoring time is assumed to be independent of the event time. However, the SDH treats the competing event as the event of interest with infinite event time. As a result, an individual with a competing event remains in the risk set (Tai, 2011). The SDH is a measure of association that reflects both the association of the treatment with the event of interest and the contribution of a competing event by actively maintaining individuals in the risk set. Thus, the use of the proportional sub-distribution hazard model depicts the rules of a certain event more reasonably. Therefore, the sample size formula for NiCTCR presented in this study is based on the proportional sub-distribution hazard model and its statistics.

**Sample size formula for NiCTCR**

The hypotheses for non-inferiority clinical trials in the presence of competing risks are as follows:

$$H_0 : \Delta \geq \Delta_0 \quad \text{vs.} \quad H_1 : \Delta < \Delta_0 ; \qquad (3)$$

i.e., $H_0 : b \geq \ln \Delta_0 \quad \text{vs.} \quad H_1 : b < \ln \Delta_0$,

where $\Delta (= \exp(b))$ is the sub-distribution hazard ratio of the event of interest. $\Delta_0 (= \exp(b_0))$ and $\Delta_1 (= \exp(b_1), \ \Delta_1 < \Delta_0)$ are the non-inferiority margin and SDH ratio under $H_1$, respectively. The Wald statistic of $b = b_0$ is $\sqrt{N}(b - b_0)$, and the estimator of $b$ is expressed as $\hat{b}$. $\sqrt{N}(b - b_0)$ is approximately normal with zero mean and variance $V(b)$, which is derived as the reciprocal of the information (Formula A.4 in the Appendix). The partial likelihood of the model (1) can be constructed as follows:

$$L(b) = \prod_{i=1}^{N} \left[ \frac{\exp(bx_i)}{\sum_{j \in RS^{SD}(T_i)} \exp(bx_j)} \right]^{I\{d_i = 1\}},$$

where $I\{CONDITION\}$ is the indicator function, which returns to 1 when the CONDITION in the braces is true; otherwise, this function returns to 0. Similar to the assignment in Latouche et al. (2004), $x_i = 0$ and $x_i = 1$ indicate that the $i$th subject is assigned to the control and experimental groups, respectively. The information is expressed as follows:

$$I_n(b) = \int_0^\infty \pi_0(t)\pi_1(t)U(t)dt ,$$

where

$$\pi_x(t) = \frac{N_x(1-F_x(t))(1-\overline{F}_x(t))}{\sum_{x=0}^{1}[N_x(1-F_x(t))(1-\overline{F}_x(t))]exp(bx)} , \text{ x = 0, 1, and}$$

$$U(t) = \sum_{x=0}^{1} p_x f_x(t)(1-\overline{F}_x(t)) .$$

$\overline{F}_x$ is the cumulative incidence function of the censoring and is assumed to be identical in the two groups, and $f_x(t)$ is the sub-density function in group $x$. The subscript $x$ indicates the corresponding statistics in group $x$. The $\pi_x(t) = N_x / N \equiv p_x$ is under the assumption $\Delta_1 = 1$; therefore, the following formula is obtained:

$$I_n(b) = p_0 p_1 \int_0^\infty U(t)dt . \qquad (4)$$

Finally, the sample size formula can be derived as follows:

$$N = \frac{(Z_{1-\beta} - Z_{\alpha/2})^2}{[\ln(\Delta_0) - \ln(\Delta_1)]^2 p_0 p_1 w} = \frac{\#E}{w} , \qquad (5)$$

where $w = \int_0^\infty U(t)dt$ implies the incidence rate of both groups, and $\#E$ implies the calculated events needed in two groups.

We assume that the survival time of the event of interest in both groups is Weibull distributed. The scale parameter of the event of interest in group $x$ is $\lambda_{x1}$, and the shape parameter denoted by $k_1$ is assumed to be identical in both groups. The survival time of the competing cause in both groups is assumed to be the same and Weibull distributed, with the respective scale and shape parameters of $\lambda_2$ and $k_2$. $\phi$ is the identical hazard rate of the exponential censoring time in both groups. $T_f$ and $R$ denote the follow-up period and accrual time, respectively. $w = \sum_{x=0}^{1} w_x$ can be calculated using the following formula:

$$w_x = \int_{T_f}^{T_f+R} \frac{q_{x1}}{R} \int_0^t k_1 \lambda_{x1} u^{k_1-1} \exp(-\lambda_{x1} u^{k_1}) \exp(-\phi u) du dt, \text{ x = 0, 1,} \qquad (6)$$

where $q_{x1}$ denotes the probability that an individual who has not experienced any event occurrence undergoes the event of interest given that individual undergoes any event; i.e.,

$$q_{x1} = \frac{\text{count of event of interest occuring in group } x}{\text{count of any event occuring in group } x}, x = 0, 1.$$

Formula (6) can be derived from the sum of two single integrals as follows:

$$w_x = \frac{q_{x1}}{R} [\int_0^{T_f} R \text{fun}(u) du + \int_{T_f}^{T_f+R} (T_f + R - u) \text{fun}(u) du], \text{ x = 0, 1,} \qquad (7)$$

where

$$\text{fun}(u) = k_1 \lambda_{x1} u^{k_1-1} \exp(-\lambda_{x1} u^{k_1}) \exp(-\phi u).$$

To plan a conservative trial, the hazard rate was based on the alternative hypothesis in Formula (3), which indicates that the cumulative incidence rates are identical in two groups, to ensure a predetermined power (Crisp, 2008).

## 3. Simulations

To verify the performance of the sample size formula of NiCTCR, 10,000 simulations of 120 parameter combinations (Table A1 in the supporting material) were conducted by using R software (V2.15.2, http://www.R-project.org). The time to failure from each cause was generated using the indirect method (Beyersmann, 2011). Fine and Gray assumed that the cumulative incidence function of the failure of interest and competing event in group $x$ follow the following model:

$$P(T_i \le t, d_i = 1 \mid x) = 1 - \{1 - q_{01}[1 - \exp(-\lambda_{01}t^{k_1})]\}^{\exp(bx)}, \ i = 1, \ldots, N,$$

$$P(T_i \le t, d_i = 2 \mid x) = (1 - q_{01})^{\exp(bx)}\{1 - \exp[-\lambda_2 t^{k_2} \exp(bx)]\}, \ i = 1, \ldots, N.$$

Consequently, $P(T_i \le t \mid d_i = D, x_i = 0)$, $i = 1, \ldots, N$, is a Weibull distribution with shape $k_D$ and scale $\lambda_{0D}$. The proportional sub-distribution hazard model (1) can be derived as follows:

$$h^{SD}(t \mid x) = \frac{q_{01}\lambda_{01}k_1 t^{k_1-1} e^{\lambda_{01}t^{k_1}}}{1 - q_{01}(1 - e^{\lambda_{01}t^{k_1}})} \exp(bx).$$

A uniformly distributed enrollment time $r_i$ and an exponential censoring time $c_i$ are generated for each individual $i$. Thus, the observed time is the minimum of $T_i$, $c_i$ and $T_f + R - r_i$.

The time to the event of interest in the control group—i.e., the baseline SDH rate—is Weibull distributed, which is conditional on the occurrence of the event of interest, with two levels of the scale parameter ($\lambda_{01} = 1, 2$) and three levels of the shape parameter ($k_1 = 0.5, 1$ and $2$). These shape parameters determine the decreasing, constant and increasing trends of the hazard curve. We assume that only one competing cause may occur and that its distribution is a Weibull distribution with scale ($\lambda_2 = 0.15$ and $0.5$) and shape ($k_2 = 0.5, 1$ and $2$) that is identical in the two groups. Different shape parameters between the failure of interest and competing failure are also considered as follows: $(k_1 = 0.5, k_2 = 1.5)$ and $(k_1 = 1.5, k_2 = 0.5)$. An exponential censoring time with 0.1 and no random censoring is considered. The detailed parameter settings are listed in Table 1 (and Table A1 in the supporting material).

The Fine-Gray proportional sub-distribution hazard model is used for modeling and estimating the confidence interval of SDH. Non-inferiority may be concluded when the upper limit of the two-sided 95% confidence interval for the SDH ratio does not exceed $\Delta_0$. Type I error and power are calculated under the presetting situation where $\Delta$ for generating data is equal to $\Delta_0$ and $\Delta_1$, respectively.

As shown in Figure 1 (and Table A1 in the supporting material), the incidence rates of the censoring and competing events were simulated from 70% to more than 20%. The three wavebands in each subgraph represent $q_{01} = 0.3, \ 0.5, \ 0.8$ in order. The dashed line represents the type I error, which is approximately 0.025 (range of 0.021 to 0.028). With the data generated under the alternative hypothesis, all of the empirical powers were approximately equal to the predetermined power of 0.8 (range of 0.791 to 0.813). Thus, the proposed sample size formula performed very well using the simulated data.

## 4. Example for NiCTCR

To illustrate the implications in terms of the sample size formulas for NiCTCR, we used the study of Schroder et al. (2009) as an example. In that study, a randomized controlled trial to evaluate the effect of early (EET, control group) versus delayed (DET, experimental group) endocrine treatment in patients with lymph node-positive prostate cancer was performed. The primary endpoint was overall survival from randomization to the day of death. The secondary endpoint was set as cancer death (CD) and non-cancer death (NCD). The Kaplan-Meier method and Cox proportional hazards model were used for the estimation, and the confidence intervals of the hazard rate of CD and NCD were used to test the non-inferiority of DET to EET with a non-inferiority margin of 1.5. In total, 234 patients were recruited to achieve 85% power and were randomly assigned at 1:1 to the EET and DET groups. In the CD comparison, NCD was treated as the competing risk.

The proposed formula was utilized for planning this trial. The required parameters were extracted from a figure (Fig. 4) in Schroder et al.'s paper (2009). The median CD time was 8.4 years for DET and 9.45 years for EET. The hazard rate was calculated as follows: $\lambda_{01} = -\ln(0.5)/9.45^{k_1}$. For low NCD, the competing cause hazard rate was derived from the survival time at a 90% survival rate, which was 5.1 years. The hazard rate for the competing risk was obtained as follows: $\lambda_2 = -\ln(0.9)/5.1^{k_2}$. The accrual interval was 12 years (1986 to 1998), whereas the follow-up time was approximately 7.5 years. The CD and NCD rates in DET were 67.1% and 23.9%, respectively. Thus, $q_{01}$ was equal to 0.737. Forty-one of 234 patients were prematurely withdrawn. Therefore, we supposed that the censoring hazard rate would equal 0.020 in the exponential censoring distribution. The sample size calculated from the presented sample size formula of NiCTCR is shown in Table 2. Several scenarios for the three shape parameters, with and without random censoring, were considered.

A total of 110 events per group was required for the presented formula to achieve 85% power. When the survival time was exponentially distributed with and without censoring, 272 and 243 patients were required, respectively. If the hazard rate decreased, the required sample size was slightly larger than that of the constant hazard because of the smaller hazard rate, whereas 288 and 269 patients per group, with and without random censoring, respectively, were necessary to increase the hazard rate survival time. When we used the sample size formula, Formula (2), for a single event or CSH in non-inferiority clinical trials (Crisp, 2008) instead of the proposed sample size formula, Formula (5), all of the sample sizes calculated were smaller than those calculated for the proportional SDH model. Single-event survival analysis overestimates the incidence proportion of the event of interest by treating the competing event as censoring. Consequently, in studies involving sub-distribution modeling, applying the sample size formula of single-event non-inferiority clinical trials may underestimate the sample size required and result in underpowered trials.

## 5. Results and discussion

In this paper, we derived a sample size formula based on the proportional sub-distribution hazard model using the Wald test in non-inferiority clinical trials, which is used for studies with a primary endpoint based on SDH. Simulations and an example showed that the presented sample size formula can correctly estimate the required sample size. The presented formula is most applicable when SDH is employed to reveal the close to "real world" relationship between the treatment and endpoints when competing risks are present, whereas the sample size formula for CSH may be used when the objective is the theoretical relationship. In many cases, apart from biological studies, learning the experimental treatment's comprehensive effects on all events may be more beneficial and attractive for an intuitive explanation (Parpia et al., 2013), and all the patients are treated in the real world where multiple events may occur. Thus, we recommend using the SDH as the primary endpoint in clinical trials when appropriate, with the presented formula to better estimate the sample size.

According to the guideline from FDA (U.S. Food and Drug Administration) for non-inferiority trials, the estimated treatment effect from the historical trials should be used to derive the non-inferiority margin for a non-inferiority trial. However, if the SDH model was not used in related studies, we suggest that the non-inferiority margin for SDH be calculated from the CSH by Lau's formula (Lau, 2009). If neither SDH nor CSH is available, a pilot study might be more appropriate.

If $\Delta_0 = 1$ and $\Delta_1 \neq 1$, the presented formula can be used for superiority testing. The body of the presented formulas and the formula of Latouche (2004) look the same. Nevertheless, the presented formula here is derived based on the Weibull distribution, where Latouche's formula is based on the exponential distribution. In addition, the calculation of parameter $\psi$ in Latouche's formula is different from that of $w$ in the denominator of the formula presented here.

One potential criticism of the competing risks application is that multiple events can be reasonably combined into a composite endpoint. For example, CD and NCD can be combined into overall survival if the detailed causes of death are not of interest and if the CD rate affected by NCD is not of concern. Nevertheless, not all events can be combined into a composite endpoint. For example, when the events defined are opposites, such as recovery vs. death for hospitalized patients with SARS (Chen, 2009), they cannot be combined into a composite endpoint. In practice, a competing risks model can reveal the underlying relationships among the studied events and variables, whereas a composite-endpoint method can simplify the analysis strategy without missing too much important information. More details are provided in recently published papers (Mell, 2010; Rauch, 2013).

The percentage of the event of interest of all of the events at the end of the follow-up interval, i.e., $q_{x1}$, was assumed to be constant in this paper. Nevertheless, the percentage can be treated as a function of the maximum follow-up time of each individual. The maximum follow-up time of each individual is not fixed because a random enrollment time exists. Therefore, the variable $q_{x1}$ in terms of the follow-up time can be adopted in Formula (6) to cover more general situations.

In this paper, the uniform accrual time was employed, i.e., the number of subjects enrolled by a Poisson distribution. In addition, the accrual rate was constant over the interval $[0, R]$. In reality, this accrual rate can be increasing or decreasing over time or can follow a non-monotone pattern. A constant rate was assumed to simplify the equations; however, other distributions have been discussed by various researchers (Maki, 2006; Crisp, 2008). For instance, other distributions can be used by incorporating the density function into Equation (6) and by replacing $1/R$.

## Appendix

A: Proof of the sample size formula in NiCTCR.

The hypothesis is as follows:

$$H_0 : b \geq \ln \Delta_0 \qquad H_1 : b < \ln \Delta_0,$$

where $\exp(b)$ is the SDH ratio $\Delta$ and $\Delta_0 (= \exp(b_0))$ is the non-inferiority margin of the SDH ratio. $\Delta_1 (= \exp(b_1))$ denotes the SDH ratio under the alternative hypothesis. The independent variable $x$ is an indicator of groups. The $x = 0$ represents the control group and $x = 1$ represents the experimental group. The risk set at time $t$ can be written as follows:

$$RS^{SD}(t) = \{i : (t \leq T_i) \cup (t \geq T_i \cap d_i \neq 1), i = 1, \ldots, N\},$$

$T_i$ denotes the time to any event of the $i$th subject and $D$ denotes the event code which is the event of interest ($D = 1$), competing event ($D = 2$) and censoring ($D = 0$). The $d_i$ is the event code of the $i$th subject. The log partial likelihood can be expressed as follows:

$$\ln L(b) = \sum_{i=1}^{N} I\{d_i = 1\} \left[ bx_i - \ln(\sum_{j \in RS^{SD}(T_i)} \exp(bx_j)) \right].$$

According to the Fine-Gray's paper (1999), the score statistic is expressed as follows:

$$s = \frac{\partial \ln L(b)}{\partial b} = \sum_{i=1}^{N} I\{d_i = 1\} \left[ x_i - \frac{\sum_{j \in RS^{SD}(T_i)} x_j \exp(bx_j)}{\sum_{j \in RS^{SD}(T_i)} \exp(bx_j)} \right].$$

The information can be derived by the following formula:

$$I_n(b) = -\frac{\partial s}{\partial b} = \sum_{i=1}^{N} I\{d_i = 1\} \left[ \frac{\left[ \sum_{j \in RS^{SD}(T_i)} x_j^2 \exp(bx_j) \right]\left[ \sum_{j \in RS^{SD}(T_i)} \exp(bx_j) \right] - \left[ \sum_{j \in RS^{SD}(T_i)} x_j \exp(bx_j) \right]^2}{\left[ \sum_{j \in RS^{SD}(T_i)} \exp(bx_j) \right]^2} \right]. \quad \text{(A.1)}.$$

If $G_i(g(x)) = \sum_{j \in RS^{SD}(T_i)} [g(x)\exp(bx_j)]$ and $x_j^2 = x_j$, then Formula A.1 can be rewritten as follows:

$$I_n(b) = \sum_{i=1}^{N} I\{d_i = 1\} \frac{G_i(x_j)G_i(1) - G_i^2(x_j)}{G_i^2(1)}, \quad \text{(A.2)}$$

where

$$G_i(1) = \sum_{j \in RS^{SD}(T_i)} \exp(bx_j) = Y_{0i} + Y_{1i}\exp(b),$$

$$G_i(x_j) = \sum_{j \in RS^{SD}(T_i)} x_j \exp(bx_j) = Y_{1i}\exp(b),$$

and $Y_{xi}$ denotes the number of patients at risk of group $x$ at $T_i$, i.e. the number of subjects within $RS^{SD}(T_i)$. Consequently, the following formula can be obtained,

$$I_n(b) = \sum_{i=1}^{N} I\{d_i = 1\} \frac{Y_{0i} Y_{1i} \exp(b)}{\left[Y_{0i} + Y_{1i} \exp(b)\right]^2} . \tag{A.3}$$

Let $F_x$ and $f_x$ be the cumulative incidence function and sub-density function of the event of interest in group $x$ and $\bar{F}_x$ be the cumulative incidence function of censoring. $N$ denotes the total number of subjects required in the two groups. $p_x$ denotes the assignment proportion of subjects to group $x$. Therefore, $Y_{xi}$ can be rewritten as follows:

$$Y_{xi} = N p_x (1 - F_x(T_i))(1 - \bar{F}_x(T_i)) .$$

Given that $U(t) = \sum_{x=0}^{1} p_x f_x(t)(1 - \bar{F}_x(t))$, the Equation (A.3) can be rewritten as follows:

$$I_n(b) = \int_0^{T_f} \frac{\prod_{x=0}^{1} [N p_x (1 - F_x(t))(1 - \bar{F}_x(t))] \exp(bx)}{\{\sum_{x=0}^{1} [N p_x (1 - F_x(t))(1 - \bar{F}_x(t))] \exp(bx)\}^2} U(t) dt ,$$

If $\pi_x(t) = \dfrac{N p_x (1 - F_x(t))(1 - \bar{F}_x(t))}{\sum_{x=0}^{1} [N p_x (1 - F_x(t))(1 - \bar{F}_x(t))] \exp(bx)}$, then

$$I_n(b) = \int_0^{T_f} \pi_0(t) \pi_1(t) U(t) dt .$$

In practice, $F_1(t) \approx F_0(t)$ is most common. In such a situation and under the alternative hypothesis, the following formula can be noted as follows:

$$\pi_x(t \mid b) \approx \frac{p_x}{p_0 + p_1 \exp(b)} .$$

Therefore, the variance is expressed as follows:

$$V(b) = (\frac{p_0 p_1}{(p_0 + p_1 \exp(b))^2} w)^{-1} \qquad \text{(A.4)},$$

where $w = \int_0^{T_f} U(t)dt$ .

The true coefficient is assumed to be known in the planning stage, hence the $b_1$ under $H_1$ is employed in the variance calculation. Therefore, the Wald test statistic is expressed as follows:

$$W = \sqrt{N}(b - b_0) \sim \text{Gussian}(0, \sqrt{V(b_1)}) .$$

The sample size formula can be derived as follows:

$$
\begin{aligned}
1 - \beta &= P(\frac{\sqrt{N}[\hat{b} - \ln(\Delta_0)]}{\sqrt{V(b_1)}} < Z_{\alpha/2} \mid H_1 : b_1 = \ln(\Delta_1) = 0) \\
&= P(\frac{\sqrt{N}[\hat{b} - \ln(\Delta_0) + \ln(\Delta_0) - \ln(\Delta_1)]}{\sqrt{V(0)}} < Z_{\alpha/2} + \frac{\sqrt{N}[\ln(\Delta_0) - \ln(\Delta_1)]}{\sqrt{V(0)}} \mid H_1 : b_1 = \ln(\Delta_1) = 0) \\
&= \Phi(Z_{\alpha/2} + \frac{\sqrt{N}[\ln(\Delta_0) - \ln(\Delta_1)]}{\sqrt{V(0)}}).
\end{aligned}
$$

After considering the inverse normal distribution function for two sides of the equation, we can obtain the following formula:

$$Z_{1-\beta} = Z_{\alpha/2} + \frac{\sqrt{N}[\ln(\Delta_0) - \ln(\Delta_1)]}{\sqrt{V(0)}} .$$

The total number of subjects can be calculated from the following formula:

$$N = \left( \frac{Z_{1-\beta} - Z_{\alpha/2}}{\ln(\Delta_0) - \ln(\Delta_1)} \right)^2 V(0) = \left( \frac{Z_\beta + Z_{\alpha/2}}{\ln(\Delta_0) - \ln(\Delta_1)} \right)^2 (p_0 p_1 w)^{-1} .$$

Under the assumption that $F_1(t) \approx F_0(t)$ and $\overline{F}_1(t) \approx \overline{F}_0(t)$ , the $U(t)$ can be expressed as $U(t) = 2 f_0(t)(1 - \overline{F}_0(t)$ . The sub-density function $f_x(t)$ can be expressed as $q_{x1} k_1 \lambda_{x1} t^{k_1 - 1} \exp(-\lambda_{x1} t^{k_1})$ (Beyersmann, 2011; Pintilie, 2006) with the Weibull survival time assumption. In this assumption, $q_{xD}$ denotes the probability that an individual who has experienced an event, undergoes the event of interest. $k_D$ denotes the shape parameter of the event $D$, which is identical across two groups. $\lambda_{xD}$ indicates the scale parameter of the event $D$ of group $x$. Let $\phi$ denote the hazard rate of censoring distribution. If no accrual interval exists, $w = \sum_{x=0}^{1} w_x$ is expressed as follows:

$$w_x = q_{x1} \int_0^{T_f} k_1 \lambda_{x1} u^{k_1-1} \exp(-\lambda_{x1} u^{k_1}) \exp(\phi u) du .$$

The same terminology found in reference (Rubinstein, 1981) required in the presence of different enrollment times considering the prolongation of the follow-up period. The incidence rate can be calculated as follows:

$$w_x = \int_{T_f}^{T_f+R} \frac{q_{x1}}{R} \int_0^t k_1 \lambda_{x1} u^{k_1-1} \exp(-\lambda_{x1} u^{k_1}) \exp(-\phi u) du dt ,$$

where $T_f$ and $R$ indicate the follow up and enrollment intervals, respectively.

**Funding**: This work was supported by the National Natural Science Foundation of

China (81673268) and the Natural Science Foundation of Guangdong Province, China

(2017A030313812).

Table 1. The parameter combination settings for sample size for NiCTCR

| Parameter | Values |
| --- | --- |
| $q_{01}$ | 0.3, 0.5, 0.8 |
| $(k_1, k_2)$ | (0.5, 0.5), (1, 1), (2, 2), (0.5, 1.5), (1.5, 0.5) |
| $\lambda_{01}$ | 1, 2 |
| $\lambda_2$ | 0.15, 0.5 |
| $\phi$ | 0, 0.1 |
| $(\Delta_0, \Delta_1)$ | (1.3, 1.0) |

Table 2. The sample size required for different parameter combinations for NiCTCR[a]

| $k_1$ | $\lambda_1$ | $\lambda_2$ | $\Delta_0$ | $\phi$ | Events[b] | $N_{CR}$[c] | $N_{SE}$[d] |
|-------|-------------|-------------|------------|--------|-----------|-------------|-------------|
| 0.5 | 0.225 | 0.047 | 1.50 | 0 | 220 | 538 | 396 |
|     |       |       |      | 0.02 | 220 | 576 | 424 |
| 1 | 0.073 | 0.021 | 1.50 | 0 | 220 | 486 | 358 |
|   |       |       |      | 0.02 | 220 | 544 | 400 |
| 2 | 0.008 | 0.004 | 1.50 | 0 | 220 | 410 | 306 |
|   |       |       |      | 0.02 | 220 | 478 | 358 |

a. $\Delta_1 = 1$, $T_f = 7.5$, $R = 12$, $\alpha = 0.05$, $q_{01} = 0.737$, power=0.85

b. Events required in two groups

c. Total sample size required by the presented formula

d. Total sample size calculated by formulas for the non-inferiority study with time-to-single-event (Crisp, 2008; Chow, 2008)

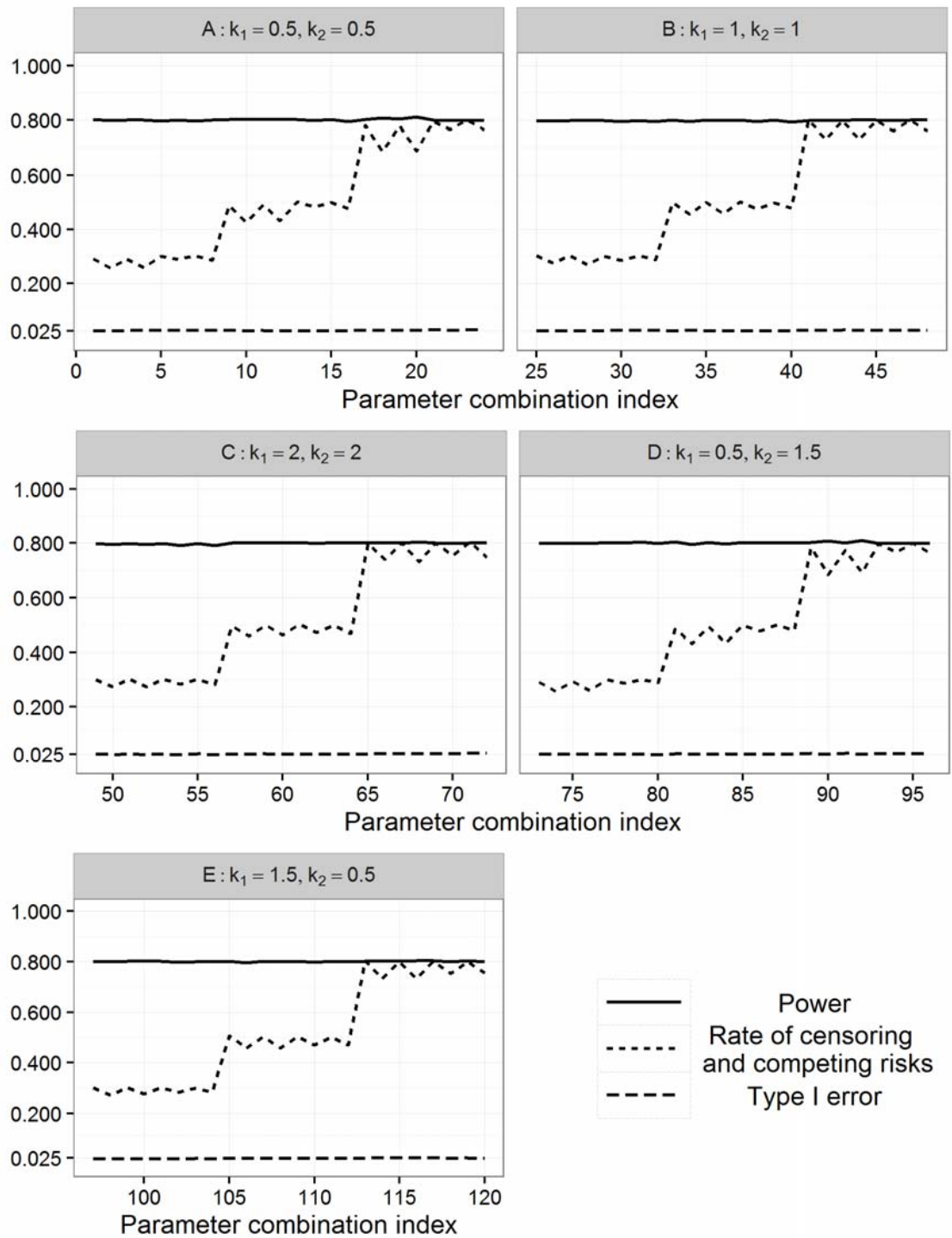

Figure 1. The power and censoring rate of the sample size formula for NiCTCR